\documentclass[useAMS,usenatbib]{mn2e}

\usepackage{graphicx}
\usepackage{times}
\title[A Search for Radio Continuum Emission]{A Search for Radio Continuum Emission\protect\\
Towards Long-period Variable Stars}
\author[Georgij M. Rudnitskij and Jessica M. Chapman]{Georgij M.
Rudnitskij$^{1}$\thanks{E-mail:
gmr@sai.msu.ru (GMR); Jessica.Chapman@csiro.au (JMC)} and Jessica M. Chapman$^{2}$\\
$^{1}$Sternberg Astronomical Institute, Moscow State University,
13 Universitetskij prospekt, Moscow, 119992 Russia\\
$^{2}$Australia Telescope National Facility, PO Box 76, Epping, NSW 2121,
Australia}

\begin{document}

\date{Accepted 2007 ... . Received 2007 ...; in original form 2007 ...}

\pagerange{\pageref{firstpage}--\pageref{lastpage}} \pubyear{2007}

\maketitle

\label{firstpage}

\begin{abstract}
We hereby report on a sensitive search for radio continuum observations
from a sample of 34 Mira and semi-regular variable stars. The
main aim of this survey was to search for thermal free-free
emission from post-shock ionised gas. Thirty-four stars were
observed at 3- and 6-cm using the Australia Telescope Compact
Array. Radio continuum emission was detected from one source
only, the symbiotic Mira R~Aqr. No continuum emission was
detected from the other sources, with three-sigma upper limits of
typically 0.3~mJy. From the upper limits to the radio flux
densities, we have found upper limits to the gas brightness
temperatures near two stellar radii at a characteristic size of
$5\times10^{13}$~cm. Upper limits to shock velocities have
been estimated using a shock model. For the 11 nearest sources
in our sample we obtain brightness temperatures below 6\,000~K 
and shock velocities below 13~km~s$^{-1}$.
For 11 out
of 14 sources with previously published detections, the radio
brightness temperatures are below 4\,000~K. For an upper limit of
4\,000~K, we estimate that the shock velocities at two stellar
radii are below 10~km\,s$^{-1}$.
\end{abstract}


\section{Introduction}

Mira variables are optically-visible long-period variable stars
with large-amplitude pulsations ($\Delta m_v>2.5^m$) and stellar
periods over 100~days. The stars have reached an advanced stage
of evolution on the asymptotic giant branch (AGB). Miras
typically have photospheric radii of $\sim 350$\,R$_{\odot}$,
stellar luminosities of $10^3\textrm{--}10^4$\,L$_{\odot}$ and
effective temperatures of $\sim$2\,500~K.

During their evolution on the AGB, Miras develop increasingly
strong stellar winds, in which mass is lost from their outer
hydrogen envelopes, with mass-loss rates $\dot{\rm M}$ above
10$^{-7}$~M$_{\odot}$\,yr$^{-1}$ and outflow velocities of
typically 8\,km\,s$^{-1}$. Such mass-loss plays a critical role
in their subsequent stellar evolution, leading eventually to the
formation of a white dwarf star and planetary nebula.

The circumstellar mass-loss of Miras and other related AGB stars
is driven by radiation pressure on solid grains, which form in the
extended hydrogen atmospheres at heights of several stellar
radii. Once formed, the grains are rapidly accelerated outwards,
while grain--gas collisions lead to an expanding circumstellar
envelope \citep{gs76}. The radiation pressure model is well
established but the physical conditions and mass-loss processes
within the grain formation zone, are poorly understood. It is
generally accepted that the high mass-loss rates are related to
the stellar pulsations and strong shock waves which propagate
outwards during each pulsation cycle, increasing the mass density
in the upper stellar atmospheres to a height where dust formation
occurs \citep*[e.g.][]{wc77,w79,m87,apv93}. Shock velocities in
the optically visible photospheres have been estimated from
hydrogen Balmer and metallic emission lines to be between 10 and
90~km\,s$^{-1}$ \citep*[e.g.][]{fwd84,gfmb85b,rw01}.

\citet{r90} and \citet{lb92} argued that at centimetre
wavelengths, a post-shock partially-ionised layer, extending to
several stellar radii, would emit optically thick thermal
free-free emission. For a Mira variable at a distance of several
hundred parsec, the expected peak flux density from such a layer
would be several milli-Janskys which is easily detectable. In
addition, thermal emission from the stellar photosphere and
non-thermal radio emission associated with stellar flare activity
might also be detectable.

So far, radio continuum emission has been detected from
approximately 15~Miras and semi-regular variables (Section 4).
\citet{rm97} reported the detection of optically-thick thermal
continuum emission from six nearby long-period variable stars.
Their VLA radio flux densities are a factor of about two larger
than expected for emission from the optical stellar photospheres
alone. \citet{rm97} provided a convincing model for their
data in which the radio emission occurs from a `radio
photosphere', of about twice the optical diameter, where the gas
temperature and density are approximately 1\,600~K and $1.5\times
10^{12}$~cm$^{-3}$ respectively. For these sources, the low radio
brightness temperatures are not consistent with strong shock
activity in the radio emitting regions. However, earlier radio
studies of several sources, including R~Aql, $o$~Cet and CW~Leo
\citep*[e.g.][]{wh73,wh77,sgk83,epr83,kbyp95} reported much
stronger radio flux densities, consistent with higher brightness
temperatures.

In this paper we describe a survey in which we used the Australia
Telescope Compact Array ({\it ATCA}) to search for radio
continuum emission at $\lambda=3$ and 6~cm from a sample of
34~Miras and semi-regular variables. The primary aim of this
study was to use radio continuum flux densities or their upper
limits as a means of investigating whether high radio brightness
temperatures, consistent with strong shock activity, may be
present in some sources.

\section{Source Selection}
\subsection{Selection Criteria}

A sample of 34 Miras and semi-regular variables was selected
using the data in the Fourth edition of the General Catalogue of
Variable Stars \citep{kh86}. Of the 34 sources, 26 are southern
Mira variables with declinations below $0^{\rm{o}}$ and visual
magnitudes at maximum brightness of $m_v<7^m$. One of the
selected Miras, R~Aqr, is a well-known symbiotic binary star. In
addition, three northern Miras, U~Ori, U~Her and R~Aql were also
observed: U~Ori and U~Her as they have well-studied OH maser
properties \citep{cc85,ccs91} and R~Aql as it has been previously
detected in radio continuum emission.

The sample also included five southern sources, R~Pic, L$_2$~Pup,
V~Hya, W~Hya and S~Pav, which are classified as semi-regular
variables but have well-defined light variations. The
semi-regular variables have smaller amplitude pulsations and
lower mass-loss rates than the Miras and might be expected to
have weaker shocks. V~Hya and W~Hya, however, have previously
been detected in radio continuum \citep{lb92,rm97}. L$_2$~Pup is
one of the closest long-period variable stars, with a
maximum-visual brightness of 2.6$^m$.

Table 1 lists properties of the observed sources. The columns of
the table are

\begin{enumerate}
\renewcommand{\theenumi}{(\arabic{enumi})}
\renewcommand{\labelenumi}{\theenumi}
\item source name
\item IRAS identification
\item variability type
\item $P$, the stellar period
\item $\langle m_v\rangle$, the {\em mean} visual maximum
\item $D$, the stellar distance
\item $\varphi$, the phase of the visual light curve
\item $S_{4.80}$, the 3$\sigma$ upper limits to the 4.8 GHz (6\,cm) flux
      density (Section 3)
\item $S_{8.64}$, the 3$\sigma$ upper limits to the 8.64 GHz (3\,cm) flux
      density (Section 3)
\item $T_b$, the upper limit to the brightness temperature of the
      radio emission, for an adopted radio size of $5 \times 10^{13}$~cm
      (Section 4.2)
\item $v_{\mathrm{s}}$, the upper limit to the shock velocity.
\end{enumerate}

\subsection{Stellar Distances}

In Table~1, the stellar periods, mean visual maxima and phases
have been taken from optical light curves provided by the
American Association of Variable Star Observers (AAVSO). For 10
of the nearest sources ($o$~Cet, R~Hor, R~Lep, L$_2$~Pup, R~Car,
W~Hya, R~Nor, R~Aql, RT~Sgr, and R~Aqr), the stellar distances
have been determined directly using trigonometric parallax
measurements recently provided by the Hipparcos astrometry
mission. For these sources, the distance errors given in Table~1
are determined from the standard errors in the stellar parallaxes
as given in the \citet{hipp97}. For one nearby source, R~Hya,
Hipparcos parallax data were not available.

For the more distant sources, with distances above $\sim$320 pc,
the Hipparcos parallax measurements are too small to provide
reliable distances. For 23~sources, the distances have been
calculated using the relationship between stellar period and
absolute visual magnitude at mean maximum given by
\citet{cf69,fhm75}. Interstellar extinction corrections were
calculated iteratively using Van Herk's model of visual
absorption \citep{vh65,cf69} with:

\begin{equation}
A_V = 0.14|\csc b|[1-\exp(-0.01D|\sin b|)],
\end{equation}

\noindent where $b$ is the galactic latitude, $D$ is the source
distance in pc and $A_V$ is the visual absorption in magnitudes.
The stellar distances were then calculated in the standard way
with

\begin{equation}
5\log D = \langle m_V\rangle -M_V+5-A_V,
\end{equation}

\noindent where $m_V$ and $M_V$ are the apparent and absolute
$V$-magnitudes at {\em mean} maximum. Mira distances calculated
in this way are accurate to typically 30\% with the uncertainties
arising largely from the intrinsic scatter in the
period--luminosity relation. Because of their irregular light
variations, larger distance errors are expected for the
semi-regular variables. For one source, V~Hya, the distance has
been determined using the near-infrared $K$-band
period-luminosity relation for oxy\-gen-rich Miras in the LMC,
given by Feast et al. (1989), with a distance modulus to the LMC
of 18.55 and average value of $K$-magnitudes given by
\citet{nl69,gsm84,fle92}.

For nine of the ten nearest sources, a comparison of the distances
obtained from the Hipparcos data with distances determined from
equations (1) and (2) shows good agreement within the likely errors.
For one source, RT~Sgr, the Hipparcos distance of 133~pc is
considerably smaller than the distance of 408~pc determined from
the $P$--$L$ relation. The reason for this discrepancy is not
clear.

\begin{tiny}
\begin{table*}
\centering
\begin{minipage}{160mm}
\caption{Stellar Properties,
Upper Limits to Radio Fluxes and Brightness Temperatures}
\footnotesize
\begin{tabular}{lcccclcccrc}
\cline{1-11}
&&&&&&&\multicolumn{4}{c}{Upper limits}\\
\cline{8-11}
~~~Star   &     IRAS     &Type& $P$  & $\langle m_{v}\rangle$ & $D^*$
&$\varphi$& $S_{4.80}$ & $S_{8.64}$ & $T_b^{**}$~~~~&
$v_{\mathrm{s}}$ \\
~~     & ~~           &    &(days)&           (mag)          & (pc) &
& \multicolumn{2}{c}{(mJy/Beam)} & (K)~~~& (km\,s$^{-1}$)\dag\\
\cline{1-11}
&&&&&&&&&&\\
S Scl  & 00128$-$3219 & M  & 363 & ~6.7 & 315
     & 0.88~     & 0.26 & 0.25 & ~13,100 &~21\\
$o$~Cet& 02168$-$0312 & M  & 332 & ~3.6 & 128$\pm$18
     & 0.58~     & 0.66 & 0.25 & ~~2,200 & ~~7\\
U Cet  & 02313$-$1322 & M  & 235 & ~7.5 & 592
     & 0.61~     & 0.30 & 0.48 & ~89\,000 & 118\\
R Hor  & 02522$-$5005 & M  & 408 & ~6.0 & 308$\pm$102
     & 0.18~     & 0.27 & 0.28 & ~14\,000 & ~22\\
R Ret  & 04330$-$6307 & M  & 278 & ~7.6 & 593
     & 0.01~     & 0.28 & 0.25 & ~46,500 & ~64\\
R Cae  & 04387$-$3819 & M  & 391 & ~7.9 & 493
     & 0.72~     & 0.30 & 0.30 & ~38,600 & ~54\\
R Pic  & 04448$-$4920 & SR & 171 & ~7.1 & 532
     & 0.20~     & 0.28 & 0.25 & ~37,400 & ~52\\
R Lep  & 04573$-$1452 & M  & 427 & ~6.8 & 250$\pm$53
     & 0.22~     & 0.22 & 0.19 & ~~6,300 & ~~13\\
U Ori  & 05528$+$2010 & M  & 368 & ~6.3 & 238
     & 0.01~     & 0.23 & 0.18 & ~~5,400 & ~11\\
V Mon  & 06202$-$0210 & M  & 341 & ~7.0 & 351
     & 0.15~     & 0.36 & 0.30 & ~19,600 & ~30\\
L$_2$ Pup & 07120$-$4433&SR& 141 & ~2.6 & ~61$\pm$5
     & 0.94~     & 0.28 & 0.26 & ~~~500 & \\
R Car  & 09309$-$6234 & M  & 309 & ~4.6 & 128$\pm$14
     & 0.12~     & 0.26 & 0.24 & ~~2,100 & ~~~7\\
S Car  & 10077$-$6118 & M  & 150 & ~5.7 & 255
     & 0.77~     & 0.24 & 0.30 & ~10,300 & ~18\\
V Hya  & 10491$-$2059 & SR & 531 & 10.0 & 380
     & 0.37:     & 0.21 & 0.21 & ~16\,000 & ~25\\
X Cen  & 11466$-$4128 & M  & 315 & ~8.0 & 613
     & 0.33~     & 0.19 & 0.22 & ~43,700 & ~60\\
U Cen  & 12307$-$5422 & M  & 220 & ~7.0 & 418
     & 0.94~     & 0.22 & 0.24 & ~22,200 & ~33\\
U Oct  & 13182$-$8357 & M  & 308 & ~7.9 & 559
     & 0.92~     & 0.96 & 0.33 & ~54,500 & ~74\\
R Hya  & 13269$-$2301 & M  & 389 & ~4.5 & 108
     & 0.63~     & 0.22 & 0.21 & ~~1,300 & ~~6\\
W Hya  & 13462$-$2807 & SR & 361 & ~6.0 & 115$\pm$14
     & 0.44~     & 0.22 & 0.23 & ~~1\,600 & ~~7\\
R Cen  & 14129$-$5940 & M  & 546 & ~6.3 & 172
     & 0.01$^p$  & 0.63 & 0.27 & ~~4,200  & ~10\\
RS Lib & 15214$-$2244 & M  & 218 & ~7.5 & 570
     & 0.12~     & 0.18 & 0.18 & ~30,900 & ~44\\
R Nor  & 15323$-$4920 & M  & 508 & ~7.2 & 195$\pm$71
     & 0.82$^p$  & 0.22 & 0.20 & ~~4\,000  & ~10\\
T Nor  & 15402$-$5449 & M  & 241 & ~7.4 & 454
     & 0.33~     & 0.27 & 0.24 & ~26,200 & ~38\\
U Her  & 16235$+$1900 & M  & 406 & ~7.5 & 399
     & 0.35~     & 0.16 & 0.17 & ~14,300 & ~23\\
RR Sco & 16534$-$3030 & M  & 281 & ~5.9 & 259
     & 0.55~     & 0.25 & 0.24 & ~~8,500 & ~15\\
RT Sco & 17001$-$3651 & M  & 449 & ~8.2 & 433
     & 0.11~     & 0.27 & 0.24 & ~23,800 & ~35\\
R Aql  & 19039$+$0809 & M  & 284 & ~6.1 & 211$\pm$53
     & 0.43~     & 0.19 & 0.17 & ~~4\,000 & ~10\\
T Pav  & 19451$-$7153 & M  & 244 & ~8.0 & 708
     & 0.51~     & 0.25 & 0.25 & ~66,300 & 89\\
S Pav  & 19510$-$5919 & M  & 381 & ~7.3 & 378
     & 0.82~     & 0.24 & 0.21 & ~15,900 & ~25\\
RR Sgr & 19528$-$2919 & SR & 336 & ~6.8 & 348
     & 0.69~     & 0.25 & 0.23 & ~14,700 & ~23\\
RU Sgr & 19552$-$4159 & M  & 241 & ~7.2 & 492
     & 0.82~     & 0.21 & 0.20 & ~25,600 & ~37\\
RT Sgr & 20144$-$3916 & M  & 306 & ~7.0 & 133$\pm$59
     & 0.86~     & 0.23 & 0.22 & ~~2,100 & ~~7\\
U Mic  & 20259$-$4035 & M  & 334 & ~8.8 & 882
     & 0.33~     & 0.20 & 0.19 & ~78,200 & 104\\
R Aqr  & 23412$-$1533 & M  & 387 & ~6.5 & 197$\pm$122
     & 0.64~     & \multicolumn{3}{c}{detected (see text)}&~~\\
&&&&&&&\multicolumn{4}{c}{} \\
\cline{1-11}
\end{tabular}
\medskip

$^*$Distances with formal errors have been determined
using Hipparcos trigonometric parallax measurements.
Other distances have been determined from
period--luminosity relations.\\
$^{**}$Brightness temperatures are calculated
for an adopted radio size of
$5 \times 10^{13}$~cm (Section 4.2).\\
\dag Shock velocities estimated using the model of \citet{fg98}.\\
$^p$primary maximum \\
\end{minipage}
\end{table*}
\end{tiny}

\section{Observations and Results}

The observations were taken on 1995 November 3--5, using the
Australia Telescope Compact Array ({\it ATCA}) with a maximum
baseline of 6~km. The {\it ATCA} is a synthesis instrument
consisting of six 22-m antennas on an east--west track, located
near Narrabri in New South Wales. The selected sources were
simultaneously observed in two frequency bands centered at
8.64~GHz (3~cm) and 4.80~GHz (6~cm) with a bandwidth in each case
of 128~MHz. Each source was observed for a total time of
approximately 75~minutes, split into five scans of 15~minutes
with the scans distributed over the sidereal time range of the
source. The data were corrected for atmospheric amplitude and
phase variations using observations of nearby strong continuum
sources, and the absolute flux density scale was calibrated using
the primary calibrator source 1934$-$638, which was taken to have
6 and 3~cm flux densities of 5.83 and 2.84~Jy, respectively.

The data reduction was carried out in a standard way using the
AIPS software package. After editing and calibrating the data,
the source visibilities were Fourier transformed using `natural
weighting'. The images were CLEANed \citep{h74,c80} and restored
with a synthesized beam of minimum FWHM 1 arcsec at 3\,cm and 2
arcsec at 6\,cm. The rms noise levels in the images were measured
empirically from the images.

Radio continuum emission was detected from one source only, the
symbiotic binary R~Aqr. This source is discussed below. For the
remaining 33 sources, columns 8 and 9 of Table~1 list the
3-$\sigma$ upper limits to the 4.80 and 8.64~GHz flux densities
which are between 0.16 and 0.96~mJy at 4.80~GHz and between 0.17
and 0.48~mJy at 8.64~GHz.

\section{Discussion}

\subsection{R Aquarii}

\begin{figure}
\begin{center}
\includegraphics[scale=.35, angle=0]{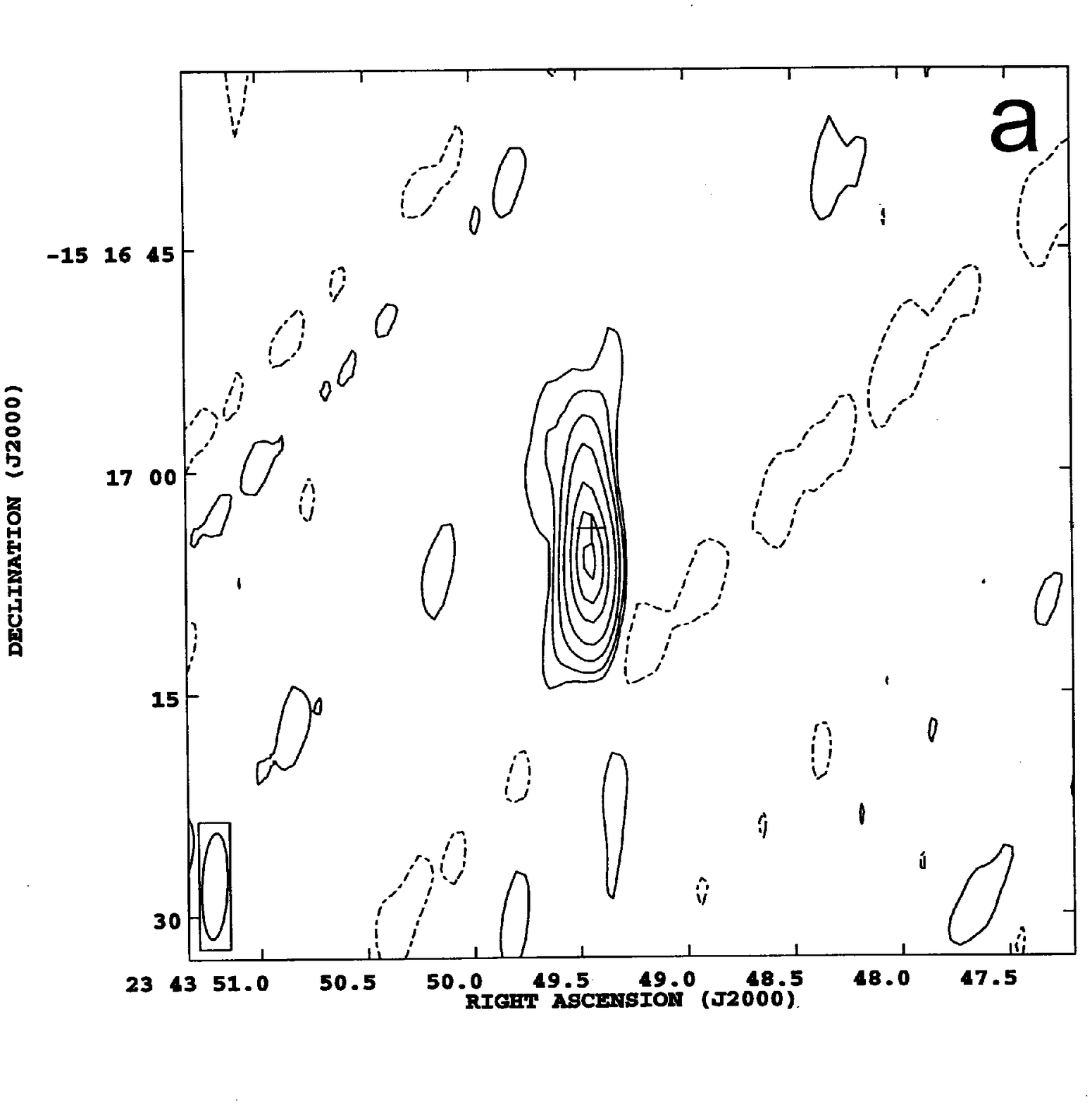}
\caption{Radio continuum image of R~Aqr at 6~cm, from data taken with
the Australia Telescope Compact Array in November 1995. The contour
levels are at $-$0.24, 0.24, 0.48, 0.96, 1.92, 3.84, 5.76 and 7.68 mJy/beam.}
\label{map_6}
\end{center}
\end{figure}

\begin{figure}
\begin{center}
\includegraphics[scale=.35, angle=0]{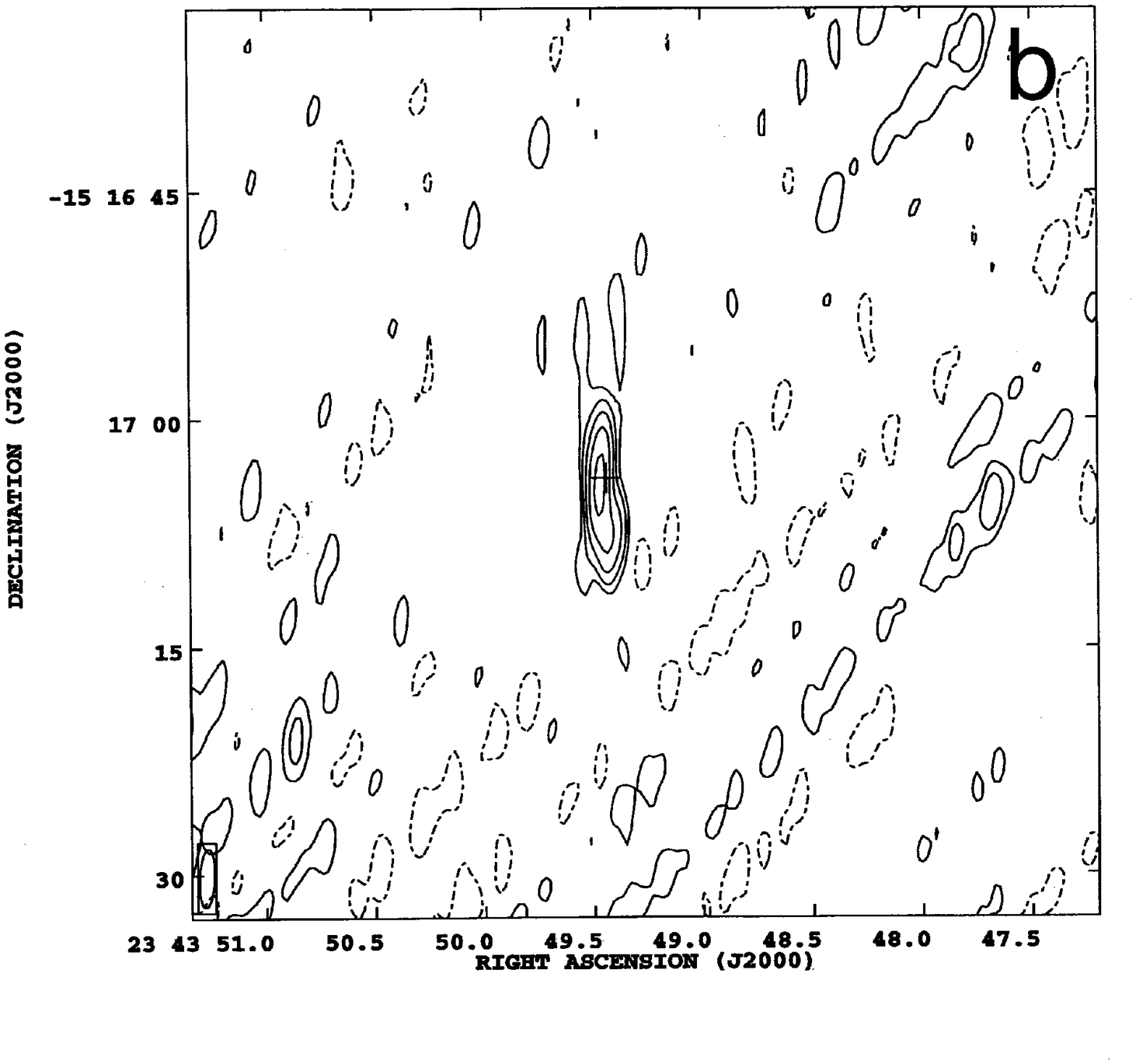}
\caption{Radio continuum image of R~Aqr at 6~cm, from data taken with
the Australia Telescope Compact Array in November 1995. The contour
levels are at $-$0.24, 0.24, 0.48, 0.96, 1.92 and 3.84~mJy/beam.}
\label{map_3}
\end{center}
\end{figure}

R Aquarii is a nearby symbiotic binary system with a cool Mira
star and a hot dwarf or sub-dwarf companion star. The latter is
evident from strongly ionised optical emission lines but is not
directly visible. The separation of the two stars has been
estimated to be $\sim15$~AU and the temperatures of the Mira and
companion stars to be $\sim 2\,800$ and 40\,000~K respectively
\citep{bvp92,hpl97}. Mass transfer between the two stars probably
occurs as the hot companion captures material from the cool
circumstellar wind of the Mira variable. Maser emission from SiO,
H$_2$O and OH molecules clearly indicates the presence of cool
dusty material associated with mass-loss from the Mira
\citep{z79,sih95}, while a complex ionised envelope, detected in
both radio continuum and optical emission, extends to at least
15\,000~AU from the binary pair.

R~Aqr also contains an astrophysical jet, first detected from
optical spectroscopy by \citep{wg80} and in radio continuum by
\citep{shkm82}. HST images have revealed a narrow, collimated and
curved jet structure which extends from within the orbit of the
binary pair to a distance of several thousand AU \citep{ph94}.

Radio continuum emission was first detected from R~Aqr in 1971
with a flux density of $\sim200$\,mJy at 85 GHz \citep{j80}.
\citet{gs74} monitored the source during 1973 and found the radio
emission to be strongly variable, on a timescale of months. The
available radio flux measurements of R Aqr (including our {\it
ATCA} result) indicate that the radio continuum emission has been
in a quiescent state since the mid-1970's. \citet{wgm81} suggested
that the initial variability may have been due to the periastron
of the companion star to R~Aqr, revolving around the Mira in a
highly eccentric orbit with $P_{\mathrm{orb}}=44$~years. This result is
supported by VLA 43-GHz observations of \citet{hpl97}. They
spatially resolved the Mira variable (detected from its SiO maser
emission) and the hot companion (detected in radio continuum) and
determined the spatial offset between the two stars to be
55$\pm$2 milliarcsec. Based on an orbital period of 44 years,
Hollis et al. estimate the semi-major axis of the orbit to be
$\sim 2.6 \times 10^{14}$~cm with a high eccentricity of $e=0.8$.

The radio continuum emission from R Aqr has previously been imaged
using several different configurations of the Very Large Array
\citep*[][and references
therein]{shkm82,khm83,khy89,hkmm85,hdy92,hpl97}. On the largest
scale, the 6-cm images of \citet{hm87,how89} show optically-thin,
thermal emission over an extended 2-arcmin region which is
associated with the optical emission nebula. On much smaller
angular scales, radio images at 2 and 6~cm show discrete
components corresponding to compact regions within the jet. The
strongest radio continuum emission occurs from a region located
close to the Mira position, which has a spectral index of $\alpha
= 0.5$ (where $S_{\nu}\propto \nu^\alpha$), consistent with
optically-thick thermal free-free emission from an outflowing
wind. Other jet components however have negative radio spectral
indices indicating non-thermal emission.

\citet{dblde95} observed R~Aqr on MERLIN at 5~GHz (De\-cem\-ber
1992) and 1.7~GHz (August 1993). In their 5-GHz images of R~Aqr
(resolution 40 milliarcsec), the jet is resolved into six,
apparently thermal components with a total flux density of
16.1$\pm$0.6~mJy.

In 1995 November the total 6-cm flux density detected with the {\it
ATCA} was 17.2~mJy (Figure~\ref{map_6}), in good agreement with the 
MERLIN value. The total
detected 3~cm flux density was 15.9~mJy (Figure~\ref{map_3}), 
corresponding to a spectral
index, averaged over the detected region, of $-0.13$.  The {\it ATCA}
radio images show the radio continuum emission from R~Aqr to be
slightly extended in a north-south direction. A comparison of the {\it
ATCA} images with the far more sensitive earlier VLA and MERLIN images
shows that radio continuum was detected from the inner jet region. The
$uv$-coverage and sensitivity of the {\it ATCA} survey data were
however not sufficient to resolve the complex structures within the
jet, or to detect the extended envelope around the binary system.

We do not discuss R~Aqr further in this paper.

\subsection{Brightness Temperature Limits}

For optically-thick radio continuum emission at centimetre
wavelengths, the gas brightness temperature, $T_b$, is related to
the radio flux density, $S_\nu$, by the Rayleigh--Jeans
approximation so that

\begin{equation}
\label{snu}
S_\nu=\frac{2\pi r^2_{\mathrm{s}}\nu^2}{D^2c^2}kT_b,
\end{equation}

\noindent where $r_{\mathrm{s}}$ is the radius of the radio-emitting region,
$\nu$ is the frequency of the radio emission, $D$ is the source
distance and $c$ is the speed of light. This may be written as

\begin{eqnarray}
\label{tb}
T_b=3.95\times10^5\left(\frac{D}{100\,{\rm pc}}\right)^2
\left(\frac{r_{\mathrm{s}}}{5\times10^{13}\,{\rm~cm}}\right)^{-2}
\times\\\nonumber
\left(\frac{\nu}{\rm GHz}\right)^{-2}
\left(\frac{S_\nu}{\rm mJy}\right)\,{\rm K}.
\end{eqnarray}

Direct optical interferometric measurements of AGB stars have
determined photospheric radii of typically
(2--3)$\times10^{13}$~cm \citep{thbf94,hst95}, while thermal
continuum emission is expected to occur from outside the optical
photosphere but within the grain condensation radii of a few
stellar radii \citep{gcm95}. From their radio data \citet{rm97}
have shown that at~cm-wavelengths the thermal radio emission from
an extended photosphere reaches unity optical depth at a radius
of approximately two stellar radii. For comparison with their
results, column 10 of Table~1 gives upper limits to the
brightness temperatures determined from equation (4) using the
stellar distances given in column 6 of Table 1 and a
`characteristic' radio size of $r_{\mathrm{s}} = 5\times
10^{13}$\,cm.

\subsection{Shock Velocity Limits}

For optically-thick thermal radio continuum emission, we can use
the observed brightness temperature, $T_b$, as a direct measure
of the post-shock electron temperature $T_e$. To obtain
`reasonable' estimates of the shock velocities we adopt a
characteristic size for the radio-emitting region of
$5\times10^{13}$~cm.

Column 11 of Table 1 gives the upper limits to the shock
velocities determined using the model of \citep{fg98} and upper
limits to the radio brightness temperatures given in column 10.
\citet{fg98} obtained a self-consistent solution of the equations
of fluid dynamics, radiation transfer and atomic kinetics,
involving partial ionisation and excitation of bound atomic
states. In Table ~2 of \citet{fg98} the dependence of the
post-shock electron temperature $T_e$ on the shock velocity
$v_{\mathrm{s}}$ is nearly linear and can be approximated as
\begin{equation}
T_e=778v_{\mathrm{s}}-3337.
\label{Tevel}
\end{equation}
Shock velocity limits have been calculated using formula
(\ref{Tevel}).

The primary result of our {\it ATCA} survey was the \emph{lack}
of any radio continuum emission detected from the 33~sources
observed (excluding the symbiotic star R~Aqr). The brightness
temperature limits given in Table~1 cover a wide range of
values, due primarily to their strong dependence on the stellar
distances.

For the six nearest sources in our sample (L$_2$~Pup, R~Hya,
W~Hya, $o$~Cet, R~Car and RT~Sgr), with distances within 140~pc,
the upper limits to the brightness temperatures are comparable to
or lower than the stellar photospheric temperatures. For these
sources, we can exclude strong 
shock activity near two stellar radii
with shock velocities above $\sim$7~km~s$^{-1}$. For the nearest source in
our sample, the semi-regular variable L$_2$~Pup, the lack of any
radio continuum detection is surprising. For this source the
upper limit to the 3-cm radio flux density is a factor of three
lower than predicted by the Reid \& Menten's model and a factor
of nearly two lower than expected for blackbody emission from an
optical photosphere of radius $3\times 10^{13}$~cm and
temperature 2\,500~K. A likely explanation is that the stellar
size is smaller than the adopted value used to estimate
brightness temperatures. Our flux density limits are consistent
with a stellar radius of $1.5\times10^{13}$~cm or smaller. For
the other five sources with distances below 140~pc, the
brightness temperature limits are consistent with the
radio-photosphere model of \citet{rm97}.

For a further five sources, R~Lep, U~Ori, R~Cen, R~Nor and R~Aql,
which are within 250~pc, the brightness temperature limits are
below 6,300~K, consistent with stellar shock velocities below
10--13~km\,s$^{-1}$. We note that these are conservative upper
limits and that actual shock velocities may be
considerably lower.

For all 15 sources in Table 1 with distances within 320~pc, we can
rule out the presence of strong stellar shocks with velocities
above 22~km\,s$^{-1}$. For the 17 sources in the sample,
with distances greater than 320~pc, our data are insufficiently
sensitive to provide useful limits to the gas temperatures or
shock velocities yield higher upper limits for $v_{\textrm{s}}$,
ranging between 23 and 118~km~s$^{-1}$.

\subsection{Comparison with Previous Radio Detections}

Many searches have been made for radio continuum emission from
long period variable stars with largely negative results
\citep[e.g.][and references therein]{wen95}. In Table~2 we
summarise previously published detections of radio continuum
emission at~cm-wavelengths from Mira and semi-regular variable
stars.

The columns of Table~2 are:

\begin{enumerate}
\renewcommand{\theenumi}{(\arabic{enumi})}
\renewcommand\labelenumi{\theenumi}
\item source name (variability type, spectral type)
\item $D$, the stellar distance. For $o$ Cet, W Hya, V Hya and R Aql,
distances are as for Table 1. For CW Leo, the distance is taken from
Crosas \& Menten (1997). For RZ Ari the distance is determined from
the Hipparcos parallax. Other distances are taken from the references
given in column 8.
\item the year of observation
\item $\nu$, the frequency (GHz)
\item $S_\nu$, the detected flux density (mJy)
\item radio telescope used
\item reference for radio detection
\item $T_b$, the radio brightness temperature calculated
assuming optically thick radio emission for a `characteristic' radius of $5\times10^{13}$~cm
(Section 4.2)
\item $v_{\mathrm{s}}$, the shock velocity calculated from the model of \citet{fg98},
using the brightness temperature given in column~8.
\end{enumerate}

\begin{table*}
\centering
\begin{minipage}{175mm}
\caption{Mira and semi-regular variable stars with previous
detections of radio continuum emission\protect\\
at~cm-wavelengths}
\footnotesize
\begin{tabular}{lcccccrcc}
\\
\hline
Star & $D^*$   & Year & $\nu$ & $S_\nu$ & Tel &
Ref & $T_b$~~~ &$v_{\mathrm{s}}$ \\
(Var, Sp)     & (pc)&      & (GHz) &  (mJy)&&
&  (K)~~~&(km~s$^{-1}$)\dag \\
\hline
\\
$o$~Cet (M,M) & ~$128\pm18$
           &1981       &~4.9 &~~0.74$\pm0.25$ & VLA     & 1 & 20,000      &30   \\
        &  &1989       &~8.4 &~~0.35          & VLA     & 2 & 3,200        &9    \\
        &  &1990/91 &~8.4 &~~0.35--0.49    & VLA     & 3 & 3,200--4,500 &9--10\\
        &  &1989       &14.9 &~~0.93          & VLA     & 2 & 2,700        &8    \\
        &  &1989/90    &14.9 &~~0.94$\pm$0.11 & VLA     & 3 & 2,700        &8    \\
        &  &1989/90    &22.4 &~~2.6$\pm$0.20  & VLA     & 3 & 3,400        &9    \\
        &  &2004/05    &~8.5 &~~0.19$^{2,3}$  & VLA     & 4 & 1,700        &7    \\
        &  &2004/05    &14.9 &~~1.06$\pm$0.37$^{3}$& VLA& 4 & 3,100        &8     \\
        &  &2004/05    &22.5 &~~1.13$^{2,3}$  & VLA     & 4 & 1,400        &6     \\
        &  &2004/05    &43.3 &~~2.52$^{2,3}$  & VLA     & 4 & ~~900        &6     \\
$\rho$~Per (SR, M) &$100\pm8$&1985 &~5   &~~0.18$\pm0.06$     & VLA  & 5
& 2,800 & 8 \\
 &   &1985 &15   &~~0.54$\pm0.08$     & VLA  & 5 & 900 & 6 \\
$\mu$ Gem (SR, M)  &$71\pm5$&1984 &~5   &~~0.18$\pm$0.05   & VLA   & 5
& 1,400 & 6 \\
R~Leo (M, M)    &$110\pm9$  &1990/91  &~8.4
&~~0.18--0.27  & VLA  & 3 & 1,200--1,800 & 6--7 \\
           &   &1989       &~8.4 &~~0.25        & VLA   & 2
& 1,700 & 7 \\
              &   &1991 &~8.4 &~~0.26$\pm$0.03     & VLA  & 6
& 1,800 &8\\
              &   &1989 &14.9 &~~0.75              & VLA  & 2
& 1,600 & 7 \\
              &   &1989/90 &14.9 &~~0.64$\pm$0.17  & VLA  & 3
& 1,400 & 6 \\
              &   &1989/90 &22.4 &~~1.47$\pm$0.20  & VLA  & 3
& 1,400 & 6 \\
CW~Leo (M, C) $^a$ &150 &1981 &~4.9 &~~0.42$\pm$0.10    & VLA  & 1
& 15,500 & 24 \\
       &    &1987 &~5   &~~0.28$\pm$0.05    & VLA  & 7
& 10,000 & 17 \\
           &    &1991 &~8.4 &~~0.77$\pm$0.03    & VLA  & 6
& 9,700 & 17 \\
           &    &1987 &15   &~~1.16$\pm$0.12    & VLA  & 7
& 4,600 & 10 \\
           &    &1985 &15   &~~1.4$\pm$0.05     & VLA  & 8
& 5,500 & 12 \\
           &    &1984 &20   &~~6.0$\pm$1.5$^2$  & OVRO & 8
& 13,300 & 22 \\
           &    &1991/93&~8.4   &~~0.67$\pm$0.03$^2$& VLA  & 9 & 8,400  &15 \\
           &    &1991/93&14.9   &~~2.14$\pm$0.18$^2$& VLA  & 9 & 8,600  &15 \\
           &    &1991/93&22.5   &~~4.20$\pm$0.46$^2$& VLA  & 9 & 7,400  &14 \\
RW~LMi (SR, C)$^b$ &230 &1991 &~8.4 &~~0.14$\pm$0.03   & VLA   & 6
& 4,100 & 10 \\
V~Hya (SR, C) & 380    &1989 &~8.4 &0.22$\pm$0.03      & VLA  & 10
& 17,800 & 27 \\
W~Hya (SR, M) & $115\pm14$ &1990/91  &~8.4
&~~0.23--0.45   & VLA  & 3 & 1,700--3,300 & 7--9 \\
         &   &1989/90 &14.9 &~~1.12$\pm$0.17 & VLA  & 3
& 2,600 & 8 \\
         &   &1989/91 &22.4
&~~2.70$\pm$0.24$^2$ & VLA  & 3 & 2,800 & 8 \\
g~Her (SR, M)    &$111\pm8$ &1989 &~8.4 &~~0.14             & VLA  & 2
& 1,000 &\\
 &    &1987 &15   &~~0.42$\pm$0.12    & VLA  & 7
& 900 & 6 \\
$\alpha^1$ Her (SR, M) &$117\pm38$ &1983/84 &~5 &~~0.83$\pm$0.06$^2$& VLA & 5
& 18,000 & 28 \\
&   &1984 &15   &~~1.71$\pm$0.10     & VLA   & 5
& 4,100 & 10 \\
R~Lyr (SR, M)   &$107\pm6$ &1987 &15   &~~0.45$\pm$0.12    & VLA   & 7
& 900  & 6 \\
R~Aql (M, M) &$211\pm53$&1982 &~2.3 &381$\pm$16      & DSN 64$^m$ & 11
& ~~~~see & text~~~~ \\
          &   &1982 &~8.4 &~40$\pm$6           & DSN 64$^m$ & 11
&''  &''\\
          &   &1970 &10.5 &~23$\rightarrow$240 & ARO 46$^m$ & 12
&''&''\\
          &   &1978 &14.9 &~~5.3$\pm$2.0       & Bonn 100$^m$ & 13
&''&''\\
          &   &1989/90 &14.9 &~~0.24$\pm0.18$  & VLA   & 3
& 1,900 & 7\\
          &   &1989/90 &22.4 &~~0.80$\pm$0.4   & VLA   & 3
& 2,800 &21\\
$\chi$~Cyg (M, S) &$135\pm11$&1989/90 &~8.4 &~~0.27$\pm$0.04 & VLA  & 3
& 2,800 &21\\
      &   &1989/90 &14.9 &~~0.79$\pm$0.20  & VLA   & 3
& 2,600 &21\\
           &   &1989/90 &22.4 &~~1.62$\pm$0.30  & VLA   & 3
& 2,300 &20\\
R~Cas (M,M)  &$160\pm13$&1989/90 &~8.4 &~~0.11$\pm$0.05  & VLA   & 3
& 1,600 & 7 \\
      &   &1989/90 &22.4 &~~0.79$\pm$0.30  & VLA   & 3
& 1,600 & 7 \\
\hline
\end{tabular}
\medskip

Notes to Table~2: \\
Other names: $^a$IRC$+$10216, AFGL 1381; $^b$CIT 6, IRC$+$30219, AFGL 1403; \\
$^1$Range of flux densities obtained from VLA monitoring data \\
$^2$Average of two or more values \\
$^3$Mira~A only\\
{\bf References:}

(1) \citet{sgk83},
(2) \citet{dlj93},
(3) \citet{rm97},
(4) \citet{mk06},\\
(5) \citet{dl86},
(6) \citet{kbyp95},
(7) \citet{dlje91},
(8) \citet{scm89},
(9) \citet{mr06},\\
(10) \citet{lb92},
(11) \citet{epr83},
(12) \citet{wh73},
(13) \citet{bk79}.
\end{minipage}
\end{table*}

Apart from R~Aqr (Section 4.1), radio continuum emission has
previously been detected from four of the sources in our sample,
V~Hya, R~Aql, $o$~Cet and W~Hya.

For the carbon star V~Hya, our three-sigma upper limit of 0.21
mJy is consistent with the 1989 VLA 8.4-GHz detection of 0.22~mJy
by \citet{lb92}.

For R Aql, strong flare activity and non-thermal emission has been
reported from several single-dish observations: \citet{wh73,wh77}
observed a rapid rise in the 10.6 GHz emission from 23 to 240~mJy
over a period of only 30 minutes. Less intense, but still
unusually strong emission from R Aql was later observed by
\citet{bk79}, while \citet{epr83} detected non-thermal emission
with a spectral index, between 2.3 and 8.4~GHz, of $-1.73$. A
possible explanation for such a steep spectral index is
non-thermal synchroton emission from electrons accelerated by a
strong stellar shock \citep{r93}. The more recent interferometric
data for R Aql however shows no evidence for excess continuum
emission. Our 8 GHz upper limit of 0.2 mJy is consistent with the
1990 VLA upper limit of 0.09 mJy \citep{rm97}. The association of
the earlier single-dish detections (which did not have accurate
radio positions) with the stellar source is unclear. We do not
however rule out the possibility that unusual and irregular flare
activity may occur from R Aql.

For W Hya, our upper limit of 0.25 mJy is slightly lower than the
mean flux density of 0.36~mJy obtained by \citet{rm97}. Their
observations partially resolved the stellar disk, with an average
diameter of 0.08 $\pm 0.02$ arcsec and a brightness temperature
of 1,500 $\pm$ 570~K .

For $o$ Cet, the three-sigma detection of \citet{sgk83} is a
marginal result. As this is inconsistent with both \citet{dlj93}
and \citet{rm97}, the published detection is probably spurious.
\citet{rm97} found marginal evidence for periodic variability in
the radio continuum emission with a mean flux density at 8.4 GHz
of 0.42 mJy and peak-to-mean ratio of $\sim$ 1.1. From a periodic
fit to the radio fluxes they showed that such variability could
be explained by a modulation of the thermal radio emission caused
by a weak pulsation-driven stellar shock propagating through the
radio-photosphere, with a shock velocity of approximately
7~km\,s$^{-1}$. Our 3-sigma upper limit at 8.6 GHz of $\sim$ 0.25
mJy is lower than the 8.4 GHz flux densities of 0.35 to 0.49~mJy
obtained by Reid \& Menten in 1990--1991. The cause of the
discrepancy between the VLA flux densities and the ATCA flux
density limit, obtained five years later, is unclear. However, it
is well-known that pulsations amplitudes and shock strengths vary
from cycle-to-cycle and some variation in the radio continuum
emission appears likely. Source variability may also occur on a
longer timescales. \citet{thb95} showed that the optical size
of $o$~Cet, measured at wavelengths between 700 and 900~nm,
increased systematically by a factor of 1.8 between 1991
September and 1993 December. The increase in optical size may
imply an episode of increased mass-loss and a build-up of the
circumstellar dust shell. Such an enhancement of circumstellar
dust density may damp the shock waves in the inner part of the
stellar envelope \citep{r97}.

The star $o$~Cet is a symbiotic binary comprising a variable red
giant (Mira~A) and an companion (possibly a white dwarf, Mira~B)
accreting material from the primary. It was resolved in VLA
observations of \citet{mk06}. The radio fluxes at 8.5, 14.9, 22.5
and 43.3~GHz were measured separately for both components at
several epochs in 2004--2005. In Table~2we list only the fluxes 
measured for the red-giant component Mira~A.

For 11 of the sources listed in Table 2 ($o$ Cet, $\rho$ Per,
$\mu$ Gem, R Leo, RW LMi, W Hya, g Her, R Lyr, R Aql, $\chi$ Cyg
and R Cas) the detected radio continuum flux densities agree well
with the radio-photosphere model of \citet{rm97}. For these
sources we estimate brightness temperatures of
1500$\rightarrow$4000~K indicating little if any excess emission.
From the shock model given in paper I, the radio brightness
temperatures indicate that only weak or moderate shock strengths
are possible with shock velocities below $\sim$ 10~km~s$^{-1}$.

As discussed by \citet{rm97}, at temperatures between $\sim$
1\,000 and 4\,000~K, free electrons occur primarily from elements
such as potassium, sodium and calcium which, although far less
abundant, have much lower ionisation potentials than hydrogen.
Their calculations show that the dominant source of the free-free
opacity occurs from interactions between free electrons from
ionised metals with neutral hydrogen atoms and molecules.

\subsection{AGB Stars with Excess Radio Continuum Emission}

Excluding R~Aql and $o$~Cet (discussed above), only three sources
in Table 2, CW Leo, V Hya and $\alpha^{1}$ Her show a strong
excess of radio continuum emission above the levels expected for
radio photospheres. We note that two of the three sources,
CW Leo and V Hya are N-type AGB carbon stars. The carbon stars
represent around eight per cent of AGB stars and have C/O ratios
enhanced surface carbon abundances due to a dredge-up of
processed material. These stars have high mass loss rates and are
considered likely to be near the end of the AGB stage of
evolution.

CW Leo (=IRC$+$10216) is an extreme carbon star (C9) and a strong
source of molecular emission, with over 50 molecular species
detected from its circumstellar envelope. From an analysis of the
CO properties, \citet{cm97} estimate the distance to be 150~pc.
For this distance, the~cm radio continuum flux densities indicate
radio brightness temperatures between $\sim$5,000 and 15,000~K.
For our model, the corresponding shock velocities are between 11
and 24~km~s$^{-1}$. For CW~Leo, the radio continuum emission
almost certainly occurs from a chromospheric region. IUE
ultraviolet spectra with ionised emission features from MgII, CII
and FeII have shown that both oxy\-gen-rich and carbon-rich AGB
stars have stellar chromospheres \citep{jl87}. \citet{ljal89}
provided a chromospheric model for N-type carbon stars. In their
model the inner chromospheres have a steep temperature gradient
and an outwards expansion velocity of around 50~km\,s$^{-1}$,
while the outer chromospheres are nearly static relative to the
stellar photospheres. The heating mechanisms in the chromospheres
of late-type stars are not well understood. However, if strong
shocks are present then some chromospheric heating is likely to
occur due to the dissipation of shock energy and heating of the
post-shock gas. \citet{mr06} monitored CW~Leo in 1991--1993 in
the radio continuum with VLA at 8.4, 14.9 and 22.5~GHz (see
Table~2). They found possible variability correlated with the
infrared phase and a spectral index very close to 2. The
variability, observed flux densities and upper limit on the
source size are consistent with the emission arising from the
stellar photosphere or a slightly larger radio photosphere.
Using the model of \citet{fg98}, we have 
estimated shock velocities that would heat
the radio photosphere to temperatures listed in Table~2; we have obtained 
$v_{\mathrm{s}}=14\textrm{--}15$~km~s$^{-1}$.

V Hya is also an extreme (C9) carbon star. CO imaging of V Hya
has shown that most of the stellar mass-loss occurs within a
high-velocity wide-angle bipolar outflow \citep{kbam96}. Such a
geometry indicates that the star is likely to be close to leaving
the AGB and may be in a short-lived common envelope binary
system. \citet{kbam96} find that along the bipolar axis there is a
strong velocity gradient with velocities decreasing outwards from
over 50~km\,s$^{-1}$ at the inner boundary to around
8~km\,s$^{-1}$ at the outer boundary. Within the bipolar cones,
multiple shocks are likely to occur as faster moving gas collides
with more slowly moving material. From this model and the above
discussion we suggest that the radio continuum emission from V
Hya, detected by \citet{lb92}, has two possible sources;
chromospheric emission from the outer atmosphere and emission
from high-velocity shocked regions located near the base of the
bipolar outflows.

Finally, $\alpha^{1}$ Her is an M5II star which belongs to a
visual binary system. For this source the 5$\rightarrow$15 GHz
spectral index of 0.8 is close to the canonical value of 0.6 for
thermal emission from an ionised wind expanding at a constant
velocity \citep*[e.g.][]{wb75}. \citet{dl86} interpret the
emission from this source as thermal emission occuring from a
chromospheric region of temperature $\sim$10,000~K and radius
$\sim$ two stellar radii. Here we note that the spectral index of
0.8 is consistent with expansion in the stellar chromosphere.

\subsection{Comparison of Radio and Optical Studies}

Considerable controversy exists on the interpretation of optical
spectra of M-type Mira and semi-regular variables. These
characteristically show hydrogen Balmer and metallic line emission
lines, which are strongest near maximum light and decline in
intensity with increasing phase towards minimum light. The
linewidths of the emission lines are also strongly phase
dependent with full widths of typically 80~km\,s$^{-1}$ near
light maximum and 40~km\,s$^{-1}$ near light minimum
\citep*[e.g.][]{gfmb85a,gfmb85b,rw01}.

From an analysis of the Balmer linewidths,
\citet{gfmb85a,gfmb85b,g88a,g88b,gld89} advocate a model in which
the double-peaked structure of the H$\alpha$ emission line
profiles observed in some M-type Miras occurs from the front and
back-side of a single spherical shock wave, with shock velocities
estimated, in most cases, to be between 40 and 70~km\,s$^{-1}$.
For S~Car, \citet{gfmb85b} estimated a maximum shock velocity of
90~km\,s$^{-1}$.

Other authors determine smaller shock velocities by arguing that
multiple shocks exist simultaneously in Mira atmospheres
\citep{w79,b88,ds90,woo95}. In this case, the H$\alpha$ emission
line profiles may have complex shapes. \citet{woo95} found three
components in the H$\alpha$ line profiles of several S-type
Miras. These were modelled as emission from different ionised
layers, associated with a `main' outward shock and an `inner'
shock caused by infalling post-main-shock gas. The shock
velocities, inferred by this model, are below 40~km\,s$^{-1}$.

From the radial velocities of optical hydrogen and metallic
emission lines, \citet{fwd84} and \citet{rw01} showed that at
maximum light the post-shock emission region has an outwards
velocity of 10--20~km\,s$^{-1}$. Similar velocities have also
been obtained from low-opacity ultraviolet emission lines
\citep{wk00}. \citet{rw01} demonstrate that the radial velocities
decrease with increasing phase, and are almost zero at minimum
light. This can be interpreted as a decrease in shock strength as
a shock wave moves outwards through the stellar atmosphere, while
the decrease in linewidth is likely to reflect a decrease in
temperature and hence Doppler broadening behind the shock front.

In the present study we have argued that the lack of any detected
radio continuum emission from the 11 nearest sources in our
sample is consistent with an upper limit to stellar shock
velocities of around 6--7~km\,s$^{-1}$ near two stellar radii.
A similar conclusion, based on the
lack of strong radio variability, was also reached by
\citet{rm97}. The lack of radio continuum detections in the
present study clearly favours low shock velocities and/or
damping of the shock energy within the inner few stellar radii.

\section{Conclusions}

We have made a sensitive search with the {\it ATCA} for radio
continuum emission at 3 and 6~cm from 34 Mira and semi-regular
variable stars. Our search provided no detections of radio
continuum emission, apart from for R~Aqr which is a well Known
symbiotic binary system. In 1995 November the radio emission from
R Aqr was in a quiescent state.

From the upper limits to the flux densities determined for the
other 33 sources, we have estimated upper limits to the gas
brightness temperatures near two stellar radii at a
characteristic size of $5\times10^{13}$~cm. Shock velocities have
been estimated using the model presented by \citet{fg98}.

For the six nearest sources in the {\it ATCA} survey, with
distances within 140~pc, we obtain brightness temperatures below
2,300~K indicating that there is no excess of radio continuum
above the levels expected from the radio-photosphere model of
\citet{rm97}. From previously published detections of centimetre
continuum emission we find that for 10 out of 13 sources the
detected flux densities indicate radio brightness temperatures
below 4,000~K. In the model of \citet{fg98} we have used, a
brightness temperature limit of 4,000~K corresponds to a shock
velocity limit of $\sim$10~km\,s$^{-1}$.

Only three long-period variable stars of those observed by other
authors, CW Leo, V Hya and $\alpha^1$ Her, show a strong excess
of radio continuum emission. In each case, the strong continuum
emission is likely to be associated with a stellar chromosphere.

\section*{Acknowledgments}

The Australia Telescope is funded by the Commonwealth of
Australia for operation as a National Facility managed by CSIRO.
G.~M.~Rudnitskij gratefully acknowledges financial support from
the Anglo-Australian Observatory and Australia Telescope National
Facility. The authors are grateful to late Dr. Janet A. Mattei
and to the American Association of Variable Star Observers (AAVSO)
for supplying the light curves for all the stars as well as to
Harm Habing and Lee Ann Willson for comments on an earlier draft
of the manuscript. This research made use of the SIMBAD database,
operated at Centre des donn\'ees astronomiques de Strasbourg
(CDS), France.

\bsp

\label{lastpage}

\end{document}